# TIME-SERIES ANALYSIS OF RECONSTRUCTED DAMA DATA

P.A. Sturrock[a,*], E. Fischbach[b], J.H. Jenkins[bc], R. Lang[b], J. Nistor[b]

[a] Center for Space Science and Astrophysics, Stanford University, Stanford, CA 94305-4060, USA
[b] Department of Physics, Purdue University, West Lafayette, IN 47907, USA
[c] School of Nuclear Engineering, Purdue University, West Lafayette, IN 47907

Corresponding author. Tel +1 6507231438; fax +1 6507234840.
Email address: sturrock@stanford.edu

Ephraim Fischbach, Ephraim@physics.purdue.edu
Jere Jenkins, jere@purdue.edu
Rafael Lang, rafael@purdue.edu
Jonathan Nistor, jnistor@purdue.edu

An analysis of DAMA data (as reconstructed from DAMA publications) confirms the presence of an annual oscillation, but with a lower significance level than that claimed by DAMA. The phase of their signal is 0.39 ± 0.02, corresponding to a peak value at about May 22, which is consistent with both the DAMA estimate and the expected phase of a dark-matter signal. However, a spectrogram analysis also shows evidence for oscillations in the frequency band 11 - 13 year$^{-1}$, that are similar to oscillations found in spectrograms formed from measurements of the decay rates of $^{36}$Cl and $^{32}$Si acquired at the Brookhaven National Laboratory (BNL). One component of these oscillations (at 11.44 year$^{-1}$) is prominent in DAMA NaI data, at the 0.2% significance level (99.8% confidence level). Analyses of BNL and other nuclear decay (specifically beta decay and K-capture) measurements point to a solar influence, either by neutrinos or by some currently unknown form of radiation. The phase of the annual oscillation in DAMA data is compatible with an influence of dark matter, and is unlikely to be attributable to a purely solar influence. We also find that annual oscillations in both $^{133}$Ba decay measurements and the Troitsk tritium-decay measurements are compatible with a cosmic influence but not with a purely solar influence. These considerations raise the possibility that DAMA measurements may somehow be influenced by a combination of solar neutrinos, cosmic neutrinos, and dark matter.

1 . Introduction

Data published by the DAMA Collaboration appear to show clear evidence for an annual modulation with a maximum at or near June 2, close to that expected for the influence of hypothetical dark matter [1-5]. Since the Collaboration has not yet released their data,[1] we have extracted their data from their publications. We present tables of the reconstructed data for the first experiment (DAMA/NaI) and the second experiment (DAMA/LIBRA) in Section 2.

The DAMA Collaboration has presented the results of power-spectrum analyses for DAMA/LIBRA data and for combined DAMA/NaI and DAMA/LIBRA data, using the Lomb-Scargle [6,7] procedure. We repeat these calculations in Section 3, with results that are consistent with those published by DAMA. However, we consider these power-spectrum calculations to be inappropriate for the DAMA

---

Footnote
1 We hope that the DAMA Collaboration will soon release their raw data—ideally the timing of all single-hit and double-hit events. The authors of this article stand ready to repeat the present analysis whenever the actual data become available.
________________________________________________________________

data for two reasons: (1) Significance estimates inferred from the power generated by a Lomb-Scargle calculation are based on the assumption that the data being analyzed have a normal distribution. We find that neither set of measurements conforms to a normal form. (2) The two datasets have quite different distributions (their standard deviations differ by a factor of 2), and it is inappropriate to combine them without taking this difference into account.

In order to obtain more reliable significance estimates, and in order to analyze data from the two experiments in combination, it is necessary to adopt a procedure that is compatible with these two requirements. One way to meet these requirements is to first apply to each dataset the rank-order normalization procedure [8], which maps measurements onto a standard normal distribution with standard deviation unity. This procedure maintains the rank order of the measurements and it would, of course, make no change in the resulting power spectrum if the initial data were actually to conform to a normal distribution. There is then no problem in analyzing the combined datasets once they have been normalized so that they both conform to exactly the same distribution.

In Section 4, we carry out this normalization and then carry out power spectrum analyses using a likelihood procedure that is equivalent to the Lomb-Scargle procedure [9,10].

In Section 5, we use a Monte-Carlo procedure (the shuffle test [11]) to obtain robust significance estimates for the oscillations that show up in the power-spectrum analyses of Section 4. The resulting significance estimate for the annual oscillation (in the combined data) is indistinguishable from that obtained in Section 4, but substantially more conservative than the 8.9 σ confidence level cited in the DAMA publications. We find that evidence in the DAMA/NaI data for an oscillation in the solar rotational band is significant at the 0.2% level (99.8% confidence level).

In order to examine the stability of the phase of the annual oscillation, we carry out power-versus-phase analyses in Section 6. We find that phases determined in this way from DAMA/NaI and DAMA/LIBRA data are virtually identical.

Since there is evidence for a rotational signal in the DAMA/NaI data but not in the DAMA/LIBRA data, we need to introduce an analysis procedure that is appropriate for the study of transient oscillations. Using such a procedure (spectrogram analysis) in Section 8, we find that the annual oscillation is very stable—significantly more stable than annual oscillations found in our recent analyses of nuclear-decay data. [12-15].

Spectrogram analysis reveals transient and/or drifting patterns in two frequency bands: 11 – 11.5 year$^{-1}$, and 12.5 – 13.0 year$^{-1}$. For comparison, we apply the same analysis procedures to data acquired at the Brookhaven National Laboratory (BNL) [16] concerning the decay rates of $^{36}$Cl and $^{32}$Si. Both of these spectrograms yield clear evidence of transient oscillations in the same two bands.

In Section 8, we study DAMA data by means of phasegrams, which are the same as spectrograms except that power is determined (in sliding time bins) as a function of phase, for the fixed annual frequency. This analysis confirms the stability of the phase (about 0.4) of the annual oscillation. We apply the same procedure to the BNL data and obtain quite different results: Analysis of the $^{36}$Cl data shows a transient annual oscillation with phase of approximately 0.7; that of the $^{32}$Si data shows a transient phase that drifts from 0.9 to 0.8. These phases are consistent with a solar origin [17].

In Section 9, we reproduce the DAMA estimate of an 8.9 σ confidence limit for the annual oscillation. We find, however, that the conditions necessary for the validity of the chi-square estimate are not met by the DAMA data.

We discuss these results in Section 10. In the Appendix, we discuss an analysis of $^{133}$Ba data acquired by the Physikalisch-Technische Bundesanstalt (PTB) Laboratory [18], and its relevance to the present analysis of the DAMA data.

## 2. DAMA Data

We have reconstructed the experimental results for the DAMA/NaI and DAMA/LIBRA experiments from data published in references [2,3]. Tables 1 and 2 present the 2 – 6 keV data for the DAMA/NaI and DAMA/LIBRA experiments, respectively.

Table 1. Measurements for the First (NaI) Experiment.

| Line Number | Date (DAMA Notation) | Date (Neutrino Days) | Date (Neutrino Years) | Residual | Day Error | Residual Error |
|---|---|---|---|---|---|---|
| 1 | 355 | 9486 | 1995.971 | -0.001 | 44.1 | 0.012 |
| 2 | 529.7 | 9660.7 | 1996.449 | 0.036 | 15.7 | 0.02 |
| 3 | 739.3 | 9870.3 | 1997.023 | -0.026 | 49.3 | 0.011 |
| 4 | 809.2 | 9940.2 | 1997.214 | 0.002 | 19.7 | 0.015 |
| 5 | 849.3 | 9980.3 | 1997.324 | 0.033 | 19.7 | 0.013 |
| 6 | 889.5 | 10020.5 | 1997.434 | 0.009 | 19.7 | 0.013 |
| 7 | 924.5 | 10055.5 | 1997.53 | -0.007 | 14.8 | 0.019 |
| 8 | 966.4 | 10097.4 | 1997.645 | -0.026 | 27.1 | 0.013 |
| 9 | 1029.3 | 10160.3 | 1997.817 | -0.038 | 34.9 | 0.013 |
| 10 | 1109.6 | 10240.6 | 1998.037 | 0 | 45.4 | 0.01 |
| 11 | 1169 | 10300 | 1998.199 | 0.015 | 14.8 | 0.019 |
| 12 | 1209.2 | 10340.2 | 1998.309 | 0.033 | 24.9 | 0.016 |
| 13 | 1265.1 | 10396.1 | 1998.462 | 0.016 | 29.7 | 0.013 |
| 14 | 1327.9 | 10458.9 | 1998.635 | 0.017 | 32.3 | 0.012 |
| 15 | 1399.6 | 10530.6 | 1998.831 | 0 | 29.7 | 0.016 |
| 16 | 1474.7 | 10605.7 | 1999.036 | -0.019 | 45.4 | 0.011 |
| 17 | 1534.1 | 10665.1 | 1999.199 | -0.004 | 14.8 | 0.018 |
| 18 | 1579.5 | 10710.5 | 1999.323 | -0.005 | 29.7 | 0.014 |
| 19 | 1644.1 | 10775.1 | 1999.5 | 0.033 | 34.9 | 0.013 |
| 20 | 1693 | 10824 | 1999.634 | 0.017 | 9.6 | 0.028 |
| 21 | 1734.9 | 10865.9 | 1999.749 | -0.009 | 28.4 | 0.019 |
| 22 | 1789.1 | 10920.1 | 1999.897 | -0.035 | 24.9 | 0.016 |
| 23 | 1859 | 10990 | 2000.088 | 0.003 | 44.1 | 0.01 |
| 24 | 1944.5 | 11075.5 | 2000.323 | 0.016 | 40.2 | 0.011 |
| 25 | 2014.4 | 11145.4 | 2000.514 | -0.001 | 29.7 | 0.015 |
| 26 | 2145.4 | 11276.4 | 2000.873 | -0.005 | 14.8 | 0.019 |
| 27 | 2175.1 | 11306.1 | 2000.954 | -0.015 | 15.7 | 0.015 |
| 28 | 2224 | 11355 | 2001.088 | -0.004 | 34.9 | 0.011 |
| 29 | 2285.2 | 11416.2 | 2001.255 | 0 | 25.8 | 0.012 |
| 30 | 2328.8 | 11459.8 | 2001.375 | 0.014 | 20.1 | 0.014 |
| 31 | 2374.2 | 11505.2 | 2001.499 | 0.009 | 24.9 | 0.013 |
| 32 | 2431.9 | 11562.9 | 2001.657 | -0.015 | 31.9 | 0.013 |
| 33 | 2494.8 | 11625.8 | 2001.829 | -0.017 | 29.7 | 0.011 |
| 34 | 2564.6 | 11695.6 | 2002.02 | -0.006 | 40.2 | 0.01 |
| 35 | 2645 | 11776 | 2002.24 | 0.004 | 40.2 | 0.011 |
| 36 | 2699.1 | 11830.1 | 2002.389 | 0.037 | 14.8 | 0.017 |
| 37 | 2734.1 | 11865.1 | 2002.484 | 0.029 | 19.7 | 0.017 |

For each table, entries in column 2 comprise the dates used by the DAMA Collaboration. However, it has been convenient to convert these dates into a format that is better suited for time-series analysis. In our studies of solar neutrinos, we have found it convenient to introduce the term "neutrino days" (column 3), comprising dates counted in days with January 1, 1970, as day 1. We also convert such measurements into "neutrino years" (column 4) as follows:

$$t(Neutrino\,Years) = 1970 + t(Neutrino\,Days)/365.2564 \ . \qquad (1)$$

This representation has the merit of being a uniformly running measure that differs only very slightly from the actual calendar date. (It avoids the leap-year problem.)

We understand that the "residuals," listed in column 5, are to be understood as measurements—in units of counts per day, per kilogram, and per keV—from which the mean value has been subtracted.

Table 2. Measurements for the Second (LIBRA) Experiment.

| Line Number | Date (DAMA Notation) | Date (Neutrino Days) | Date (Neutrino Years) | Residual | Day Error | Residual Error |
|---|---|---|---|---|---|---|
| 1 | 3195.2 | 12326.2 | 2003.747 | -0.006 | 28.4 | 0.006 |
| 2 | 3250 | 12381 | 2003.897 | -0.011 | 24.8 | 0.008 |
| 3 | 3300 | 12431 | 2004.034 | -0.008 | 24.8 | 0.007 |
| 4 | 3350.1 | 12481.1 | 2004.171 | 0.008 | 24.8 | 0.007 |
| 5 | 3395.6 | 12526.6 | 2004.295 | 0.01 | 20 | 0.008 |
| 6 | 3430.2 | 12561.2 | 2004.39 | 0.008 | 14.8 | 0.009 |
| 7 | 3460.2 | 12591.2 | 2004.472 | 0 | 14.6 | 0.009 |
| 8 | 3523 | 12654 | 2004.644 | -0.006 | 48 | 0.006 |
| 9 | 3584.8 | 12715.8 | 2004.814 | -0.016 | 14.8 | 0.01 |
| 10 | 3614.9 | 12745.9 | 2004.896 | -0.011 | 14.8 | 0.01 |
| 11 | 3654.9 | 12785.9 | 2005.005 | -0.005 | 24.8 | 0.007 |
| 12 | 3710.4 | 12841.4 | 2005.157 | 0.001 | 30 | 0.007 |
| 13 | 3795 | 12926 | 2005.389 | 0.011 | 24.8 | 0.009 |
| 14 | 3835.1 | 12966.1 | 2005.499 | 0.017 | 14.8 | 0.01 |
| 15 | 3865.1 | 12996.1 | 2005.581 | 0.005 | 14.1 | 0.011 |
| 16 | 3897.9 | 13028.9 | 2005.671 | 0.004 | 18.9 | 0.009 |
| 17 | 3940.6 | 13071.6 | 2005.788 | 0.013 | 24.6 | 0.008 |
| 18 | 3979.8 | 13110.8 | 2005.895 | -0.017 | 15 | 0.009 |
| 19 | 4019.8 | 13150.8 | 2006.004 | -0.01 | 24.8 | 0.006 |
| 20 | 4069.8 | 13200.8 | 2006.141 | -0.002 | 24.8 | 0.007 |
| 21 | 4115.3 | 13246.3 | 2006.266 | 0.004 | 20 | 0.007 |
| 22 | 4159.9 | 13290.9 | 2006.388 | 0.013 | 24.8 | 0.007 |
| 23 | 4210 | 13341 | 2006.525 | 0.002 | 24.8 | 0.007 |
| 24 | 4267.3 | 13398.3 | 2006.682 | -0.007 | 32.1 | 0.008 |
| 25 | 4320.1 | 13451.1 | 2006.826 | -0.005 | 20 | 0.007 |
| 26 | 4360.1 | 13491.1 | 2006.936 | -0.016 | 20 | 0.007 |
| 27 | 4405.6 | 13536.6 | 2007.061 | -0.002 | 25 | 0.006 |
| 28 | 4460.2 | 13591.2 | 2007.21 | 0.006 | 30 | 0.006 |
| 29 | 4514.8 | 13645.8 | 2007.36 | 0.014 | 24.8 | 0.007 |
| 30 | 4560.3 | 13691.3 | 2007.484 | 0.003 | 20 | 0.008 |
| 31 | 4613.1 | 13744.1 | 2007.629 | -0.004 | 32.3 | 0.005 |
| 32 | 4674.9 | 13805.9 | 2007.798 | -0.004 | 30 | 0.005 |
| 33 | 4725 | 13856 | 2007.935 | -0.01 | 20 | 0.007 |
| 34 | 4780.5 | 13911.5 | 2008.087 | 0.001 | 35.3 | 0.005 |
| 35 | 4840.5 | 13971.5 | 2008.251 | 0.008 | 25 | 0.006 |
| 36 | 4890.6 | 14021.6 | 2008.388 | 0.014 | 25 | 0.006 |
| 37 | 4935.2 | 14066.2 | 2008.51 | 0.004 | 20 | 0.007 |
| 38 | 4973.4 | 14104.4 | 2008.615 | -0.007 | 18.4 | 0.009 |
| 39 | 5079.8 | 14210.8 | 2008.907 | -0.011 | 20 | 0.006 |
| 40 | 5135.4 | 14266.4 | 2009.059 | -0.007 | 35.3 | 0.004 |
| 41 | 5205.4 | 14336.4 | 2009.25 | -0.001 | 35.3 | 0.004 |
| 42 | 5275.5 | 14406.5 | 2009.442 | 0.01 | 35.3 | 0.004 |
| 43 | 5333.7 | 14464.7 | 2009.602 | 0.009 | 23.7 | 0.005 |

**3. Lomb-Scargle Analysis**

We now attempt to reproduce the power spectra computed by the DAMA Collaboration, concerning which there is some confusion. The caption of their Figure 2 of ref. [3] reads *Power spectrum of the measured single-hit residuals ...calculated according to Refs. [41,42], including also the treatment of the experimental errors and of the time binning.* The DAMA refs. [41,42] correspond to our refs [6,7], which present what is now known as the Lomb-Scargle procedure. However, the confusion is that the Lomb-Scargle procedure does *not* take account of *either* the experimental errors *or* the time binning (i.e. the start and stop times of each measurement).

To attempt to resolve this issue, we have applied the Lomb-Scargle procedure to our reconstructed DAMA dataset. For comparison with Figure 2 of the DAMA article, we first apply the Lomb-Scargle procedure to the DAMA/LIBRA data. This yields the power spectrum shown in Figure 1. For comparison with our investigations of anomalous beta-decay data, we choose to cover the frequency range 0 – 20 year$^{-1}$ (whereas the DAMA article covers only the range 0 – 2.9 year$^{-1}$). In our analysis, the peak power at 1 year$^{-1}$ has the value S = 13.3. The value we find in the left-hand panel of Figure 2 of the DAMA article is 14.2. These values are close enough to suggest that our reconstruction of the DAMA data is not far off. It also confirms that the DAMA analysis does in fact use a simple Lomb-Scargle procedure that does not take account of either the experimental errors or the time binning (in disagreement with the caption to the DAMA figure).

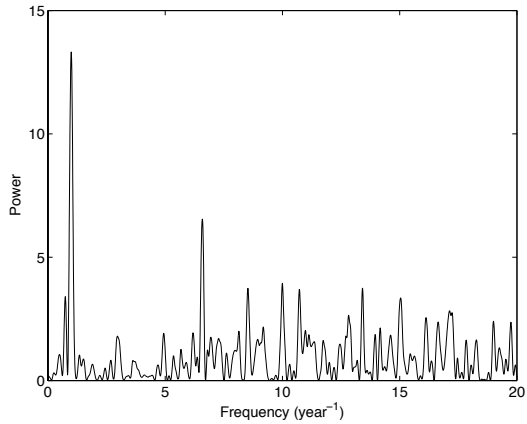

Figure 1. Power spectrum analysis of DAMA/LIBRA data, using the Lomb-Scargle procedure.

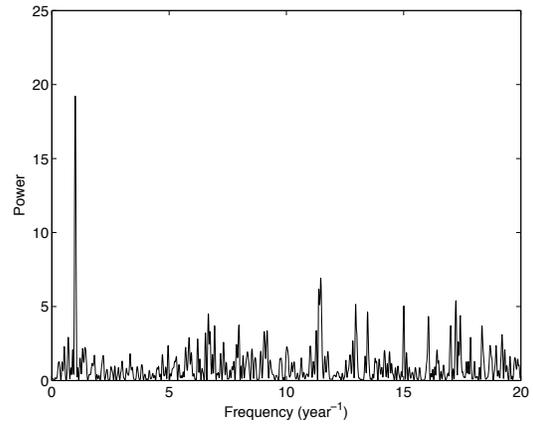

Figure 2. Power spectrum analysis of the combined DAMA/NaI and DAMA/LIBRA data, using the Lomb-Scargle procedure

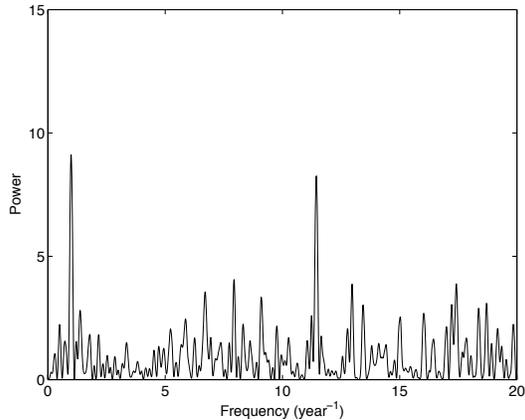

Figure 3. Power spectrum analysis of the DAMA/NaI data, using the Lomb-Scargle procedure.

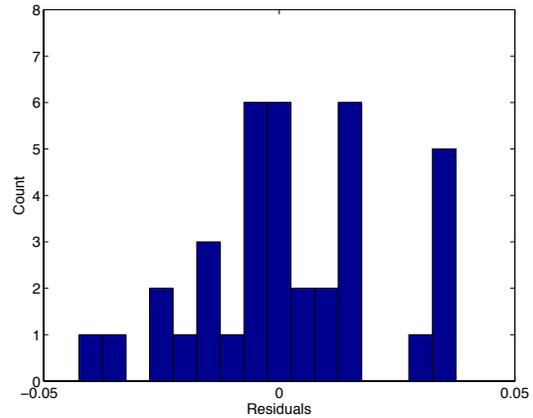

Figure 4. Histogram of residual measurements from the DAMA/NaI Experiment

For comparison with the right-hand panel of Figure 2 of the DAMA article, we have applied the Lomb-Scargle procedure to a simple concatenation of the DAMA/NaI data and the DAMA/LIBRA data. The result is shown in Figure 2, in which the power of the annual oscillation is found to be 19.2. The relevant figure in the DAMA article has a similar peak with power 21.4.

We draw attention to the fact that Figure 2 shows a peak at frequency 11.47 year$^{-1}$ with power 6.9, since this falls within the search band (10 – 15 year$^{-1}$) we have adopted for possible evidence of the influence of solar rotation [14,15].

The peak at about 6.7 year$^{-1}$, which is especially prominent in DAMA/LIBRA data, does not have an obvious relationship to known solar periodicities, but examination of additional datasets may shed light on this feature.

For completeness, we have also applied the Lomb-Scargle procedure to the DAMA/NaI data, with the result shown in Figure 3. The peak at 1 year$^{-1}$ now has a power of only 9.1. It is interesting to note that the peak at 11.45 year$^{-1}$ is almost as strong (with power 8.3), which raises the possibility that the DAMA experiment may have been subject to an intermittent solar influence.

**4. Power-Spectrum Analysis using Rank-Order Normalization**

We have carried out Lomb-Scargle analyses in Section 3 since that is a standard procedure for power-spectrum analysis of data with irregular spacing, and for comparison with results reported by the DAMA Collaboration. However, the Lomb-Scargle procedure has two shortcomings for application to DAMA data.

The first and most serious problem is that the DAMA/NaI and DAMA/LIBRA experiments are quite different, so that the resulting time series are quite different. Figures 4 and 5 show the histograms of measurements from the two experiments. We note specifically that the standard deviations are quite different— 0.0194 for the DAMA/NaI residuals, but only 0.0091 for the DAMA/LIBRA residuals, a ratio of 2:1.

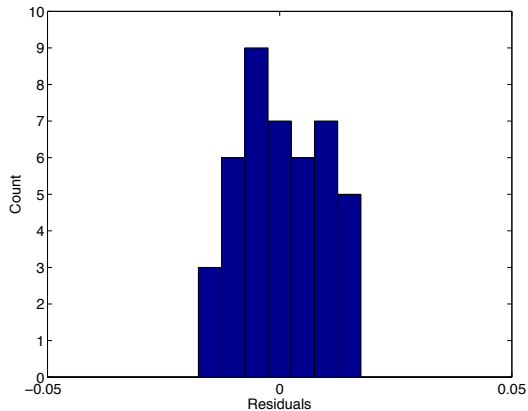

Figure 5. Histogram of residual measurements from the DAMA/LIBRA Experiment.

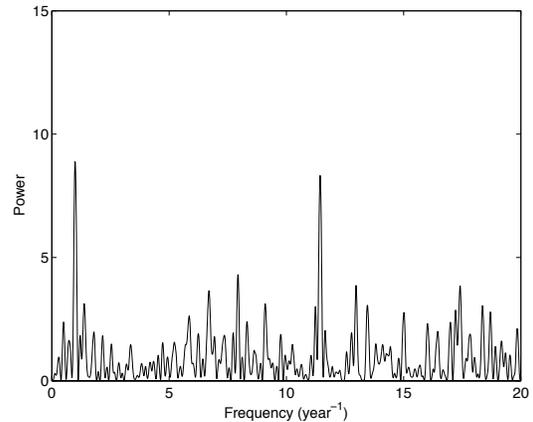

Figure 6. Power spectrum analysis of the DAMA/NaI data, using the rank-order renormalization procedure, followed by a likelihood analysis.

The second problem is that the Lomb-Scargle procedure is designed for application to time-series in which the background contribution (the "noise") has a normal distribution. As we see from Figures 4 and 5, neither histogram resembles a normal distribution.

For these two reasons, we have chosen to re-analyze the DAMA data using the rank-order normalization (*rono*) procedure [8]. For any set of measurements, this operation retains the exact rank-order of the measurements, but maps them onto a normal distribution. As a result, we can expect the resulting power spectrum to have an exponential distribution (apart from any true oscillations that may be present). It has the additional advantage that there is now no problem in concatenating two different sets of measurements (that may have quite different distributions), since the *rono* operation converts them into a standard (normal) form.

Using these procedures, we have generated power spectra for the DAMA/NaI and DAMA/LIBRA datasets, as shown in Figures 6 and 7, respectively, and for the combined dataset, as shown in Figure 8.

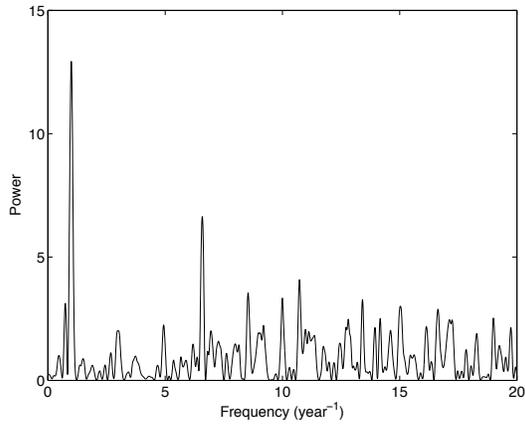

Figure 7. Power spectrum analysis of the DAMA/LIBRA data, using the rank-order renormalization procedure, followed by a likelihood analysis.

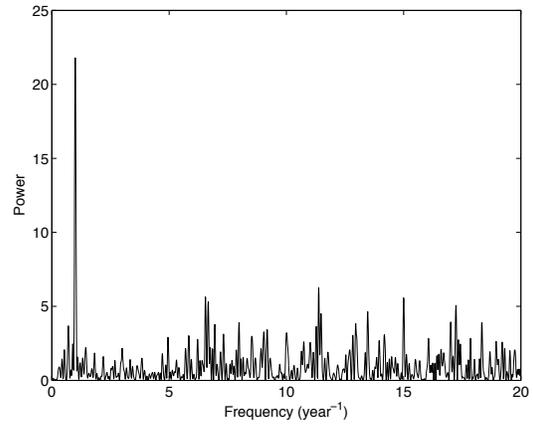

Figure 8. Power spectrum analysis (by a likelihood procedure) of the combined DAMA/NaI and DAMA/LIBRA data, after applying the rank-order renormalization procedure to each dataset separately prior to concatenation.

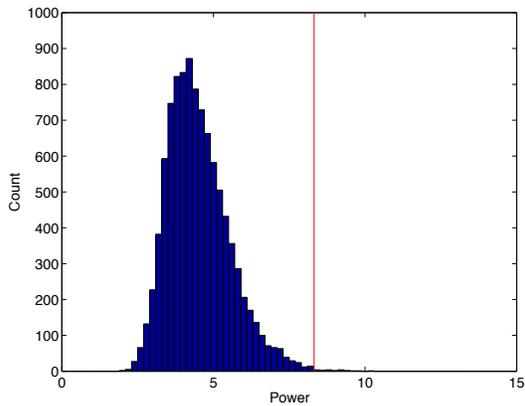

Figure 9. Histogram of the maximum power in the frequency range 10 – 15 year$^{-1}$ in power spectra formed form 10,000 shuffle simulations of the DAMA/NaI data.

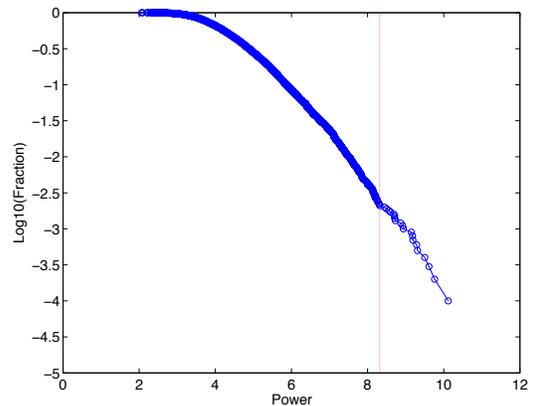

Figure 10. Logarithmic display of the maximum power in the frequency range 10 – 15 year$^{-1}$ in power spectra formed form 10,000 shuffle simulations of the DAMA/NaI data. OPnly 20 simulations have a power as large as or larger than the actual maximum power in that range (8.32), indicating that the oscillation is significant at the 0.2% level.

## 5. Monte-Carlo Simulations using the Shuffle Procedure

We need to obtain significance estimates for the oscillations found in our power-spectrum analyses. One can derive estimates from the power of a peak at a specified frequency. For instance, the probability of finding a peak of power $S$ or more at a specified frequency fro power-spectrum analysis of data that are independent and conform to a normal distribution is given by $e^{-S}$ [7]. We see from Figure 8 that the power of the annual oscillation in the combined DAMA/NaI and DAMA/LIBRA datasets is 21.80. Hence we expect that the probability of finding a peak with this power or more at the annual frequency to be about $3 \times 10^{-10}$. However, we can obtain a more robust estimate of the significance level by generating Monte Carlo simulations of the data. A simple and convenient procedure for generating simulations is the shuffle procedure in which one retains the actual times and the actual values of measurements, but randomly re-assigns values to times [11].

We first apply this procedure to the DAMA/NaI dataset. We have generated 10,000 simulations of the data by the shuffle procedure. For each simulation, we determine the maximum power in the range 10 – 15 year$^{-1}$, which is the band we adopt for the possible influence of solar rotation [14,15]. The result is shown in histogram form in Figure 9 and in a logarithmic display in Figure 10. We find that only 20 simulations out of 10,000 have a power as large as or larger than the actual maximum power (S = 8.32 at $\nu$ = 11.44 year$^{-1}$) in that band. We may infer that this oscillation is significant at the 0.2% level (99.8% confidence level).

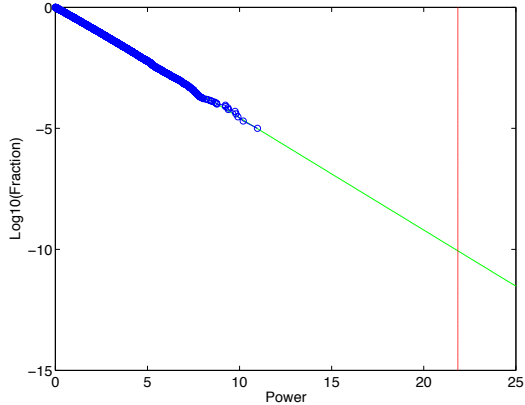

Figure 11. Logarithmic display of the maximum power at 1 year$^{-1}$ in power spectra formed form 100,000 shuffle simulations of the concatenated DAMA/NaI and DAMA/LIBRA datasets. Power spectra were computed by a likelihood .A projection of the resulting curve indicates that one would expect to find by chance only one simulation out of about $10^{10}$ with power as large as or larger than the actual power (21.85). A significance level of $10^{-10}$ corresponds to 6.3 $\sigma$ .

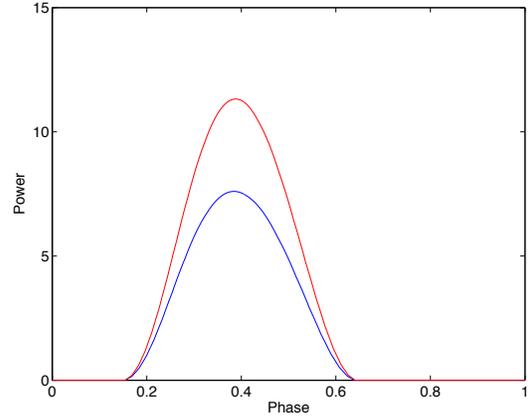

Figure 12. Power spectra computed by a likelihood procedure, as a function of phase, for DAMA/NaI (blue) and DAMA/LIBRA (red).

We have also applied this procedure to obtain a significance estimate of the annual oscillation. We now examine the combined data in the form of a concatenation of the DAMA/NaI and DAMA/LIBRA datasets, each of which has been normalized by means of the *rono* operation. We have carried out 100,000 Monte Carlo simulations by means of the shuffle procedure, shuffling the two datasets independently before concatenating them. Figure 11 presents a logarithmic display of the result. By extrapolating the resulting curve, we find that the probability of finding a power of 21.85 or more by chance is only about $10^{-10}$. This estimate is very close to the one we inferred from the power.

A significance level of $10^{-10}$ corresponds to 6.3 $\sigma$, which is impressive, but is considerably more conservative than the estimate of 8.9 $\sigma$ that we find in the DAMA publications. We discuss this discrepancy in Section 9.

**6. Phase Analysis**

We now carry out power-spectrum analyses by the same procedure as in Section 5, except that we now fix the frequency (1.00 year$^{-1}$) and determine the power for each possible value of the phase. By the term "phase" we refer to the phase of the year at which the modulation is a maximum.

We show the results in Figure 12. For DAMA/NaI, the modulation is found to peak at phase 0.38 ± 0.03. For DAMA/LIBRA, the modulation is found to peak at phase 0.39 ± 0.03. The 1-$\sigma$ error estimates are taken to be the phase shifts at which the power has dropped by 0.5 below the peak value. Clearly, the phase of the annual modulation is very stable. When we merge DAMA/NaI and DAMA/LIBRA, we find that the modulation peaks at 0.39 ± 0.02, with peak power 21.80. This peak phase converts to day 142 ± 7, or May 22 ± 7.

**7. Spectrogram Analysis**

We have seen in Section 5 that, although the annual modulation is very stable, the rest of the power spectra differ significantly. It is of course possible that the difference is due to systematic or other changes. However, the first Monte Carlo test in Section 6 indicates that the oscillation at 11.44 year$^{-1}$ in the DAMA/NaI data is quite significant. For these reasons, we should consider the possibility that, apart from the stable annual modulation, the DAMA experiments may be subject to transient influences.

The fact that we have found a modulation in the solar-rotation frequency band raises the possibility that the DAMA experiments may somehow be related to beta-decay experiments which have been found to exhibit solar-related oscillations that are typically transient in nature. We have found that power-spectrum analysis is of limited usefulness in the study of transient oscillatory phenomena— we obtain better insight by supplementing power-spectrum analyses with spectrograms. We shall focus on two frequency bands— one that includes the annual oscillation, and another that covers the rotational band (10 – 15 year$^{-1}$).

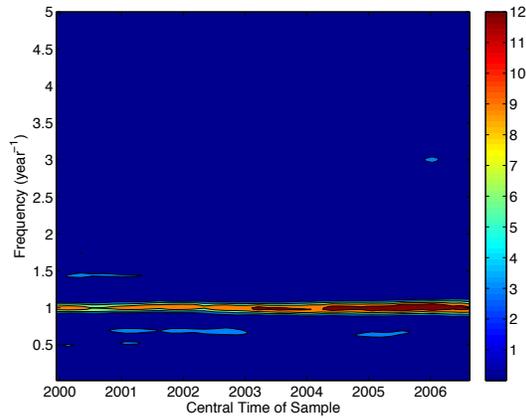

Figure 13. Spectrogram formed from combined DAMA/NaI and DAMA/LIBRA data by power-spectrum analysis of sequences of 40 measurements. We see that there is a strong and stable oscillation at 1.00 year$^{-1}$.

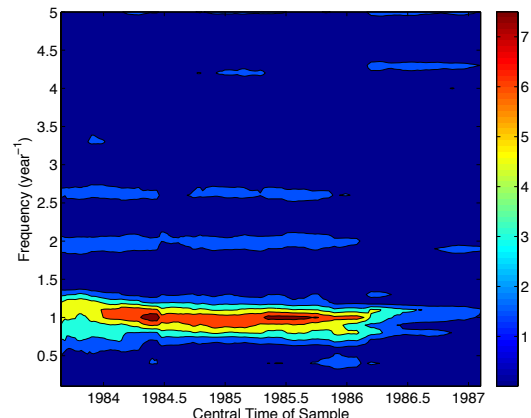

Figure 14. Spectrogram formed from BNL $^{36}$Cl data, showing a strong but intermittent oscillation at 1.00 year$^{-1}$.

We form spectrograms by generating power-spectra from a sequence of short sections of measurements. The spectrogram shown in Figure 13 was generated from power-spectrum analysis of sections that each comprise 40 measurements. The individual power spectra are generated by the same procedures as used in Section 5. As expected from the power spectra shown in Figures 6, 7, and 8, the annual oscillation is seen to be very stable. The power is weaker before 2003 than it is after that date, due presumably to the improvement of DAMA/LIBRA over DAMA/NaI. There is little to be seen in this spectrogram other than the annual oscillation.

It is interesting to compare spectrograms formed from DAMA data with spectrograms formed from measurements of anomalous beta decay. We show in Figures 14 and 15 spectrograms formed from $^{36}$Cl and $^{32}$Si decay data obtained by Alburger, Harbottle and Norton in experiments carried out as the Brookhaven National Laboratory (BNL) [16]. We see that there is a clear annual oscillation in the $^{36}$Cl data, although the power varies significantly with time. The spectrogram formed from $^{32}$Si data also shows evidence for oscillations in the neighborhood of 1 year$^{-1}$, but the frequency and power are highly variable.

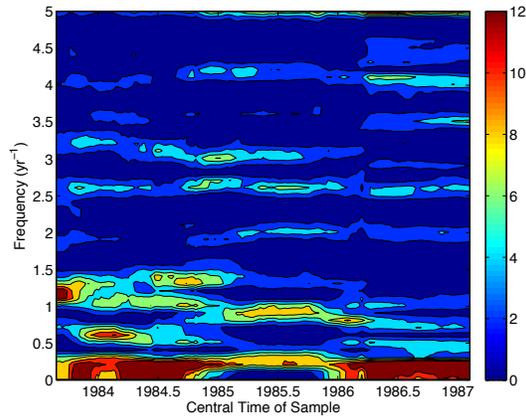

Figure 15. Spectrogram formed from BNL $^{32}$Si data, showing only a sporadic oscillation at 1.00 year$^{-1}$.

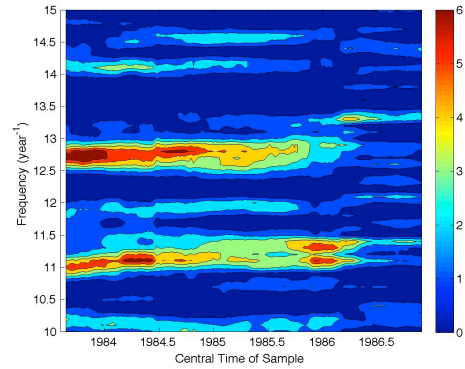

Figure 16. Spectrogram formed from BNL $^{36}$Cl data, showing evidence of two intermittent oscillations in the search band for rotational effects, one in the band 11.0 to 11.5 year$^{-1}$, and the other in the band 12.5 to 13.0 year$^{-1}$.

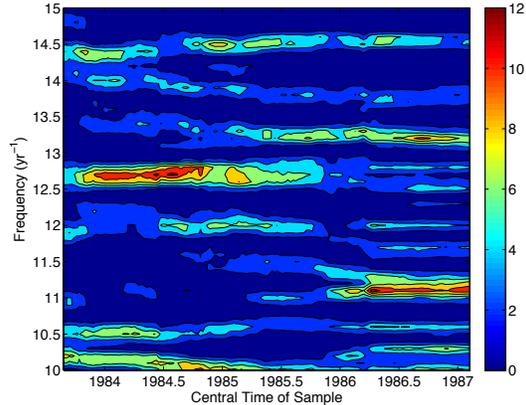

Figure 17. Spectrogram formed from BNL $^{32}$Si data, showing evidence of two intermittent oscillations in the search band for rotational effects, one in the band 11.0 to 11.5 year$^{-1}$, and the other in the band 12.5 to 13.0 year$^{-1}$.

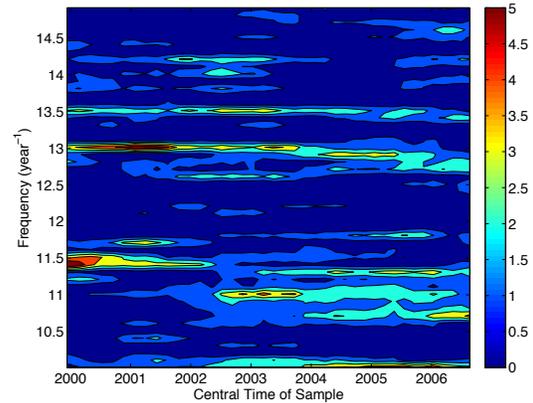

Figure 18. Spectrogram formed from combined DAMA/NaI and DAMA/LIBRA data showing evidence of two oscillations in the search band for rotational effects, one in the band 11.0 to 11.5 year$^{-1}$, and the other in the band 12.5 to 13.0 year$^{-1}$.

We next examine the rotational band 10 – 15 year$^{-1}$. Figures 16 and 17 show spectrograms formed from $^{36}$Cl and $^{32}$Si data, respectively. We see clear evidence for oscillations in two sub-bands: 11 to 11.5 year$^{-1}$ and 12.5 to 13 year$^{-1}$. Figure 18 shows the spectrogram formed from DAMA data for the same frequency band 10 – 15 year$^{-1}$. We see evidence for oscillations in the two sub-bands that are prominent in the BNL data. The oscillation at about 11.5 year$^{-1}$, which is very prominent in Figure 6, is the strongest feature in this figure, but is of short duration. This is of course the peak (at 11.47 year$^{-1}$) that is prominent in the power spectrum shown in Figure 6. We discuss these results further in Section 10.

## 8. Phasegram Analysis

We have also carried out phasegram analyses, in which we focus on the annual oscillation and examine the power as a function of date and phase. The phasegram generated in this way from DAMA data is shown in Figure 19. We see that the phase is consistently in the neighborhood of 0.4, but is somewhat more stable in the DAMA/LIBRA interval than in the DAMA/NaI interval.

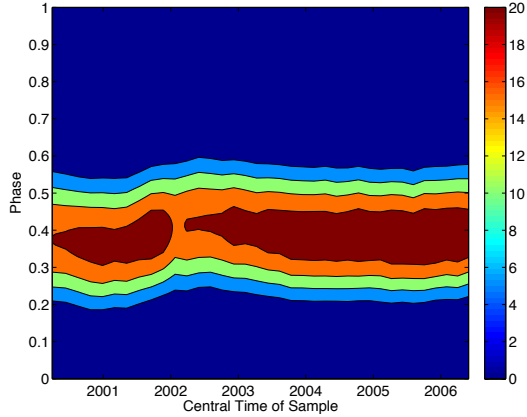

Figure 19. Phasegram formed from combined DAMA/NaI and DAMA/LIBRA data showing that the oscillation at 1.00 year$^{-1}$ has a fairly steady phase at about 0.4 year$^1$.

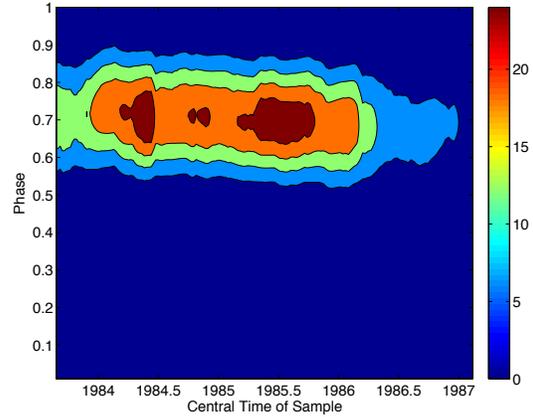

Figure 20. Phasegram formed from BNL $^{36}$Cl data showing that the intermittent oscillation at 1.00 year$^{-1}$ has a phase of about 0.7 year.

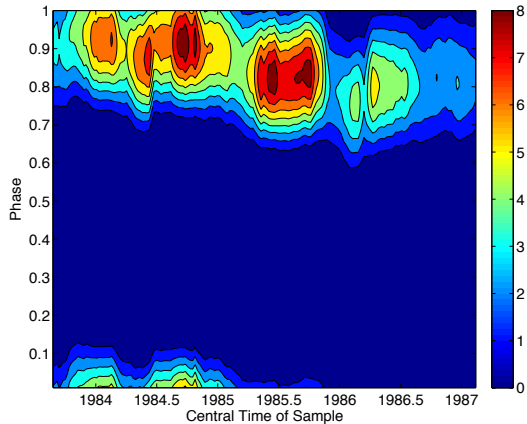

Figure 21. Phasegram formed from BNL $^{32}$Si data showing that the intermittent oscillation at 1.00 year$^{-1}$ has a phase that drifts from 0.9 to 0.8 year.

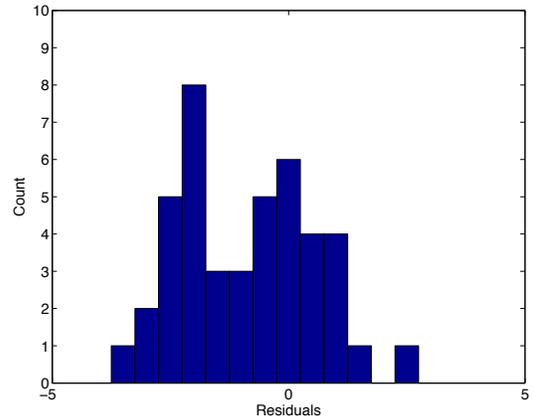

Figure 22. Histogram of residual measurements less the maximum-likelihood fit, divided by the residual error estimates, for the combined DAMA/NaI and DAMA/LIBRA data.

On the other hand, we see from Figures 20 and 21 that the phasegrams generated from the two BNL datasets are quite different. For the $^{36}$Cl data, the phase is fairly stable at about 0.7. For the $^{32}$Si data, the phase drifts from about 0.9 to about 0.8.

We discuss the phase issue in Section 10.

### 9. DAMA Significance Estimates

The key result presented by the DAMA Collaboration is the presence of an annual modulation in their data. The crucial issue is the significance level of this result. We here review the claim made by the DAMA Collaboration.

In one of their most recent articles [4], they claim *model-independent evidence of the presence of DM particles in the galactic halo at 8.9 σ C.L. on the basis of the investigated DM signature*. For this result, the authors refer to their ref. [17], which is our ref. [3]. In that article, the relevant significance levels are listed in Table 3. The last item in that table is for the combined DAMA/NaI and DAMA/LIBRA data, for the energy band 2 – 6 keV. This item lists a confidence limit of 8.8 σ, which corresponds to a P-Value of 5.0 × 10$^{-19}$. However, that item also lists a chi-square value of 64.7 for 79 degrees of freedom, but these figures lead to a P-Value of 0.88. As a result, it is not clear how the Confidence Level estimate of 8.8 σ has been obtained.

It seems clear, however, that DAMA has obtained its Confidence-Level estimates from chi-square calculations, as in ref. [5]. We have attempted to reproduce the DAMA estimate by forming a chi-square statistic as follows:

$$X = \sum_{n=1}^{M} \frac{(r_n - r_{ML})^2}{re_n^2}, \qquad (2)$$

where $r_n$ and $re_n$ are the residuals and the errors of the residuals, as listed in Tables 1 and 2, and $r_{ML}$ is the maximum-likelihood value that minimizes $X$. Combining all DAMA data, we find that the chi-square value $X$ has the value 216.2. For 79 degrees of freedom, this leads P = 6 × 10$^{-15}$ as the probability that the annual modulation may have arisen by chance. This value of $P$ corresponds to an 8.8 σ or 8.9 σ effect, which is the value claimed by DAMA.

We now look for an explanation of the discrepancy between this confidence estimate and the value (6.9 σ) that we found in Sections 4 and 5. We note that the standard formula for the significance of a chi-square value is based on the assumption that, on the null hypothesis, each measurement is distributed according to a normal distribution with standard deviation given by the error estimate. In order to check this assumption, we examine the distribution of the terms appearing in the summation of Equation (2), as shown in Figure 22. Since this figure does not resemble a normal distribution (it is highly asymmetric), it is clear that one should not expect the standard chi-square calculation to give an accurate confidence estimate.

**10. Discussion**

The principal concern about the DAMA experiment has been the conflict between the apparent detection of dark matter by DAMA and the non-detection by other dark-matter experiments, notably CDMS [19] and XENON100 [20]. In this section, we explore the possibility that the influence on the detector may not be that of dark matter.

Our analysis has confirmed that the DAMA experiment detects an influence with a precise annual frequency and a very well defined phase, which we estimate to be 0.39 ± 0.02. This converts to May 22 ± 7, which is close to the DAMA estimate of May 26 ± 7, and consistent with the estimate (June 2) predicted by the annual modulation signature of dark matter [21,22].

Our most reliable significance estimate was obtained by a Monte Carlo analysis using the shuffle procedure. We find that there is a probability of only 10$^{-10}$ of finding by chance an annual modulation as strong as or stronger than that which we find in the combined DAMA/NaI and DAMA/LIBRA data. This converts to a 6.3 σ confidence limit, which is more conservative than the 8.9 σ confidence limit proposed by the DAMA Collaboration.

We draw attention to evidence for a solar influence on the DAMA experiment. We found from Monte Carlo simulations that there is a probability of only 0.002 (99.8% confidence level) of finding by chance a modulation in the rotational search band of 10 – 15 year$^{-1}$ that has a power as large as or larger than the actual peak power (8.32 at 11.44 year$^{-1}$) in that band in the power spectrum formed from DAMA/NaI data. This leads us to consider the possibility that the DAMA results may somehow be analogous to the behavior of experiments that provide evidence for variability in some nuclear decay processes. As summarized in references [12-15 and 17], experiments at several laboratories (including the Brookhaven National Laboratory [16] and the Physikalisch-Technische Bundesanstalt [18]) have yielded evidence that some nuclides exhibit variability in beta-decay and/or electron-capture processes. This possibility has more recently been noted independently by Pradler et al. [23].

The possibility that DAMA may be seeing an annual modulation arising from a radioactive source in their experiment has been previously considered [24], and is the subject of an article now in preparation [25].

We have found that the anomalous variability of certain decay processes is typically transient in nature. In order to determine whether oscillations in DAMA data are transient, we examined spectrograms formed from those data in Section 7, with the interesting result that, although the annual oscillation is very stable, the same is not true of oscillations with frequencies in the rotational band. There appear to be two sub-bands— 11.0 to 11.5 year$^{-1}$ and 12.5 to 13.0 year$^{-1}$ —where we have found evidence of variability not only in the DAMA data, but also in measurements of the decay rates of $^{36}$Cl and $^{32}$Si acquired at the Brookhaven National Laboratory (BNL).

We see from the spectrograms of Figures 14 and 15 that the BNL data also contain evidence for annual oscillations, but these oscillations are transient in nature, whereas (as we see in Figure 13) the annual oscillation in DAMA data is very stable. We also find that the phases of the annual oscillations in the BNL data are quite different from the phase of the annual oscillation in the DAMA data. We see from Figures 20 and 21 that the phases of the annual oscillations in the BNL data are in the range 0.7 to 1.0, whereas that of the DAMA oscillation is approximately 0.4. The BNL phases are consistent with a model we have developed to explain the range of phases found in most decay experiments (approximately 0.7 to 1.0 and 1.0 to 0.2) [17]. The phase of the annual oscillation in the DAMA data is not compatible with that model.

These considerations suggest that the DAMA experiment may be responding to two distinct influences: one solar, and one non-solar. The solar influence is manifested by the oscillations in the rotational search band. The non-solar influence is manifested by the annual oscillation. We should note, however, that there may be a solar contribution to the annual oscillation, which could influence the phase and could lead to some variability.

The precise mechanism whereby the Sun influences certain decay rates is unknown, but the leading candidate seems to be neutrinos, or possibly neutrino-like particles that we may refer to as "neutrellos." Neutrinos are produced in enormous numbers in the solar core, and they appear to travel freely through the outer layers of the Sun, except that they are subject to oscillations by the MSW (Mikheyev, Smirnov, Wolfenstein) [26,27] and RSFP (Resonant Spin Flavor Precession) [28-31] mechanisms, which can lead to variability in any one flavor of neutrinos. The mechanism by which neutrinos might influence beta-decays and/or electron-capture events is unknown.

Baurov [32] and Parkhomov [33] have suggested that nuclear decays may be influenced by cosmic neutrinos, and Falkenberg has proposed that radioactive decays are influenced by solar neutrinos and "other sources" of neutrinos [34]. Additionally, Jenkins and Fischbach [35] and Fischbach et al. [36]. discuss the possibility of detecting relic cosmic neutrinos via radioactive decays. These authors note that the estimated flux of cosmic neutrinos ($\sim 1 \times 10^{10}$ cm s$^{-1}$) is similar to variations that have been detected in the flux of solar neutrinos (flux $\sim 6 \times 10^{10}$ cm s$^{-1}$). These two sources of neutrinos are believed to have quite different energies, but the significance of this energy difference for nuclear decays is unknown.

We find supporting evidence—discussed in the Appendix—for this possibility in measurements of the decay rate of $^{133}$Ba, acquired by the Physikalisch-Technische Bundesanstalt (PTB) [18]. We find that $^{133}$Ba decay rates show a strong annual oscillation with phase in the neighborhood of 0.4. Lang is currently studying the decay of $^{133}$Ba, and he also raises the possibility that $^{133}$Ba is influenced primarily by cosmic neutrinos. We also find, from an analysis of data generously made available by the Troitsk Collaboration, that there is evidence of an annual modulation with phase close to 0.4 in tritium decay measurements. Lobashev et al. [37] and Stephenson et al. [38,39] have discussed the possibility that Troitsk measurements may be influenced by cosmic neutrinos. In an article now in preparation [25], we explore the possibility that the decay rate of $^{40}$K, which is known to be present in the DAMA NaI detectors, may be influenced not only by solar neutrinos and cosmic neutrinos, but also by dark matter.

We are indebted to Dr Heinrich Schrader for kindly making available the PTB measurements concerning [133]Ba, and to Dr. Nikita Titov and Dr Vladimir Lobashev for kindly providing us with data acquired by means of the Troitsk experiment.

**Appendix. [133]Ba Data Analysis**

The Physikalisch-Technische Bundesanstalt (PTB) has measured the decay rates of 8 nuclides over a period of about 30 years [18]. Herr Schrader has kindly made their data available to us. We here summarize our analysis of the data concerning [133]Ba.

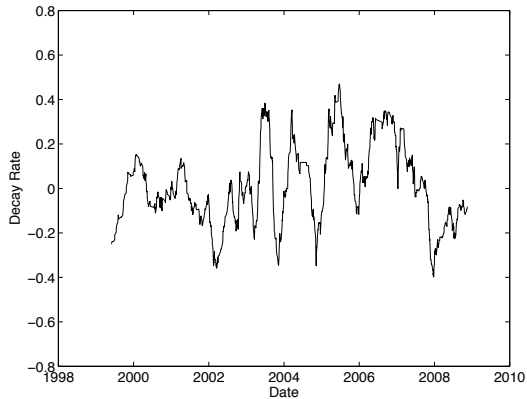

Figure 23. 51-point running means of the normalized PTB decay-rate measurements for [138]Ba.

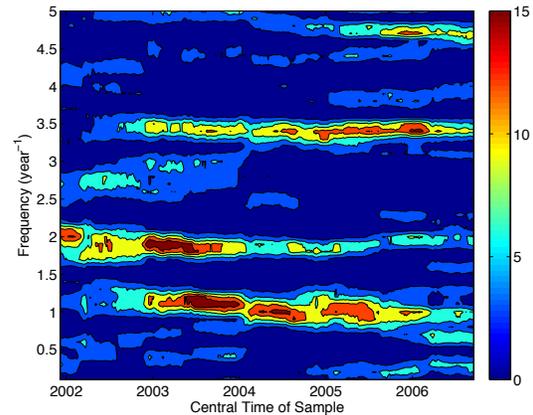

Figure 24. Spectrogram for frequency range $0 - 5$ year$^{-1}$ formed from PBS [133]Ba measurements, showing a strong annual modulation.

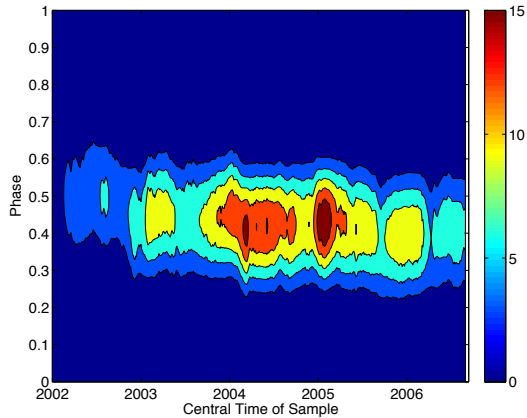

Figure 25. Phasegram for the annual modulation of PBS [133]Ba measurements, showing a peak with phase about 0.43 (June 7).

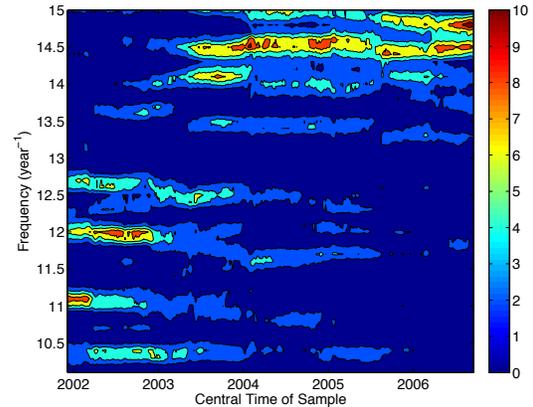

Figure 26. Spectrogram for the frequency range $10 - 15$ year$^{-1}$ formed from PBS [133]Ba measurements. We see evidence of oscillations with frequencies 11.0 year$^{-1}$, 12 year$^{-1}$ and 12.5 year$^{-1}$, similar to those found in BNL [36]Cl and [32]Si data (Figures 16 and 17).

The dataset comprises 1091 measurements made over the time interval 1999.412 to 2008.871. Figure 23 shows 51-point running means of the measurements after they have been normalized to remove the exponential decay. There is an obvious but sporadic annual modulation, which becomes more obvious in the spectrogram shown in Figure 24. This plot also shows evidence of an oscillation at about 2 year$^{-1}$, which is suggestive of a non-sinusoidal waveform of the annual modulation.

We show in Figure 25 a phasegram generated for the annual modulation. A likelihood analysis of the data indicates that the mean phase is 0.43 ± 0.04, which is consistent with the phase of the annual modulation found in the DAMA data (and consistent with what might be expected of dark matter).

Figure 26 shows a spectrogram formed for the frequency range 10 – 15 year$^{-1}$. It is interesting that the figure shows evidence of three oscillations similar to those (at about 11 year$^{-1}$, 12 year$^{-1}$, and 12.5 year$^{-1}$) found in the BNL $^{36}$Cl and $^{32}$Si data (Figures 16 and 17). In order to assess the significance of these oscillations, we have carried out 1,000 shuffle simulations of the $^{133}$Ba data and formed the same spectrogram for each simulation, noting the maximum power in the frequency range 11 – 13 year$^{-1}$. The result is shown in histogram form in Figure 27.

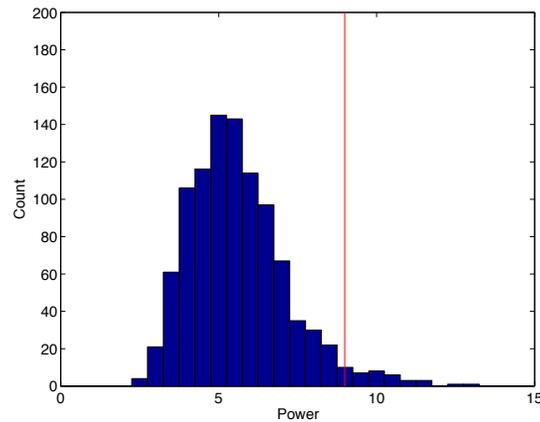

Figure 27. Histogram display of the results of 1,000 shuffle simulations of the $^{133}$Ba data. Only 36 of 1,000 have a peak power in the frequency band 11 – 13 year$^{-1}$ equal to or larger than the actual peak power, 8.99.

We find that only 36 of the simulations have a power equal to or larger than the actual peak power (8.99) in that band, indicating that the rotational oscillations in the $^{133}$Ba data are significant at the 4% level (96% confidence level).